\documentclass[fleqn,usenatbib]{mnras}

\usepackage{newtxtext,newtxmath}

\usepackage[T1]{fontenc}

\DeclareRobustCommand{\VAN}[3]{#2}
\let\VANthebibliography\thebibliography
\def\thebibliography{\DeclareRobustCommand{\VAN}[3]{##3}\VANthebibliography}

\usepackage{graphicx}	
\usepackage{amsmath,ulem}	
\usepackage{natbib}
\usepackage{breqn}
\usepackage{amsmath}
\usepackage{siunitx}
\usepackage[dvipsnames]{xcolor}
\usepackage{cancel} 

\usepackage{bigints}
\usepackage{tikz}
\PassOptionsToPackage{dvipsnames}{xcolor}
\definecolor{webgreen}{rgb}{0,.5,0}

\def\be{\begin{equation}}
\def\ee{\end{equation}}
\def\ba{\begin{eqnarray}}
\def\ea{\end{eqnarray}}

\def\12{{1\over 2}}

\def\msun{M_\odot}

\def\etal{{\it et~al.~}}
\def\ltgt{$\; \buildrel < \over \geq \;$}
\def\gtlt{\lower.5ex\hbox{\ltgt}}
\def\ltsima{$\; \buildrel < \over \sim \;$}
\def\simlt{\lower.5ex\hbox{\ltsima}}
\def\gtsima{$\; \buildrel > \over \sim \;$}
\def\simgt{\lower.5ex\hbox{\gtsima}}

\definecolor{arsenic}{rgb}{0.23, 0.27, 0.29}

\def\blue{\textcolor{blue}}
\definecolor{falured}{rgb}{0.5, 0.09, 0.09}
\definecolor{ao(english)}{rgb}{0.0, 0.5, 0.0}

\title[Dust-free starburst galaxies at redshifts $z>10$]{Dust-free starburst galaxies at redshifts $z>10$}

\author[Nath et al]
{
 Biman B. Nath$^{1}$, Evgenii O. Vasiliev$^{2}$, Sergey A. Drozdov$^{2}$, Yuri A. Shchekinov$^{1}$  \\
\footnotesize \it $^{1}$Raman Research Institute, Sadashiva Nagar, Bangalore 560080, India;\\
\footnotesize \it $^{2}$Lebedev Physical Institute, Russian Academy of Sciences, 53 Leninsky Avenue, Moscow 119991;
}
\date{Accepted XXX. Received YYY; in original form ZZZ}

\pubyear{2015}

\begin{document}
\label{firstpage}
\pagerange{\pageref{firstpage}--\pageref{lastpage}}
\maketitle

\begin{abstract}
\setlength{\leftskip}{10em}
One of the most distant galaxies GN-z11 was formed when the Universe was $\le$ 400 Myr old, and it displays a burst-like star formation rate $\sim 25~\msun$ yr$^{-1}$ with a metallicity $Z\sim 0.2\pm 0.1Z_\odot$. It resembles $z=2-3$ galaxies (at ``cosmic noon") except for the fact that the measured reddening $E(B-V)=0.01\pm 0.01$ indicates the presence of little or no dust. This marked absence of dust hints towards violent dynamical events that destroy or evacuate dust along with gas out of the galaxy on a relatively short time scale and make it transparent. We apply a 3D numerical model to infer possible physical characteristics of these events. We demonstrate that the energetics of the observed star formation rate is sufficient to tear apart the dusty veil on time scales of $20-25$ Myr. This can explain the apparent lack of evolution of UV luminosity function of galaxies between and $z\ge 10$ and $z\sim 7$, by compensating for the lower galaxy masses at higher redshift by the absence of dust. We show,  however, that this is a temporary phenomenon and soon after the last of the supernovae explosions have taken place, the expanding shell shrinks and obscures the galaxy on time scales of $\approx 5-8$ Myr.  
\end{abstract}

\begin{keywords}
\setlength{\leftskip}{10em}
galaxies: high-redshift, formation, evolution, ISM
\end{keywords}


\section{Introduction} 
Only one spectroscopically confirmed very high-redshift ($z\geq 10$) galaxy is known to date, dubbed GN-z11, at redshift $z=11.09^{+0.08}_{-0.12}$, when the age of the Universe was $\simeq 400$ Myr. GN-z11 has been identified as a bright ($M_{uv}\approx -22$ mag), massive (stellar mass $M_\ast\simeq 10^9~\msun$) Lyman break galaxy in observations with the {\it Hubble Space Telscope}/WFC3/IR and {\it Spitzer}/IRAC instrument \citep{Oesch2016}. Near-infrared observations of [CIII]1907, CIII]1909 doublet and OIII]1666 lines with Keck I telescope allowed the measurement of the redshift more precisely as $z=10.957_{-0.001}^{+0.001}$ \citep{Jiang2020}. Furthermore, \citet{Harikane2022a} reported the photometric identification of two even brighter ($M_{uv}\approx -24$ mag) galaxy candidates at $z\simeq 13$. Very recently, the {\it James Webb Space Telescope} (JWST) has found two more bright and massive galaxies with $M_{uv}\simeq -21$ mag and $M_\ast\simeq 10^{9}~\msun$: GL-z13 (photometric $z=13.1^{+0.8}_{-0.7}$) and GL-z11 ($z=10.9^{+0.5}_{-0.4}$) with its CEERS and GLASS programes \citep{Castellano2022, Naidu2022}. These recently discovered bright galaxies challenge the conventional scenario of UV luminosity functions \citep{Naidu2022}, since the total number of the currently known $z>10$ bright galaxies would require 10 times larger area for being detected within the standard paradigm of structure formation. Overall, the early JWST results and the ALMA REBELS galaxy survey suggest an unexpected abundance of $z>10$ $M_{UV}\simeq -21$ galaxies. The properties of these  high-z galaxies have been discussed in \citep{Harikane2022a,Naidu2022}, although more data are needed to firmly conclude about their nature. 

The case of GN-z11 is intriguing becasue its large stellar mass and metallicity ($\sim 0.1\, Z_\odot$) are at odds with the observed absence of dust attenuation, $A_{uv}<0.2$ mag 
\citep{Oesch2016}.
Although relatively high metal content ($\sim 0.05Z_\odot$), 
has also been found in a $z\sim 7.2$ galaxy \citep{Inoue2016}, it is reasonable to think that GN-z11 is not an exception, but a common phenomenon among the first galaxies. Moreover, inspired by recent finding of exceptionally bright galaxies in the $z\simgt 10$ universe, \citet{Ferrara2022} have considered the evolution of the UV luminosity function (LF) from $z\sim 14$ to $z\sim 7$ and hypothesized that $z>11$ galaxies have negligible amount of dust \citep[see also][]{Mason2022}. \citet{Fiore2022} have invoked radiatively driven wind (through dust grains) in this context. 
\citet{Arata2019} had also deduced that stellar feedback could make high-$z$ galaxies intermittently UV-transparent, although the SFR in their simulated galaxies had a much smaller value than found in GN-z11.

In this paper we discuss supernovae driven wind  that can explain the puzzle, and also include the effect of dust destruction behind shock waves. In addition we discuss a rather surprising aftermath of such a wind, to the effect that a dust-free condition may prevail for a limited duration of time.
This paper is organized as follows. Section \ref{sc} briefly describes the proposed scenario. In Section \ref{mod} we describe the  model set-up for the simulation, and present our results in \ref{sec:num}, followed by a discussion on the implications.

\section{The scenario}\label{sc}  
 
\subsection{Estimates: mass budget and optical depth}\label{sec:optdept}

All recently found $z>10$ galaxies are recognized to be star-forming or a short-term starburst Lyman break galaxies (LBG) \citep{Oesch2016,Harikane2022a,Naidu2022}. Using the Kennicutt-Schmidt \citep{Kennicutt2012} law for  
the local universe ($\dot\Sigma_\ast\approx 10^{-2} \Sigma_g$, $\dot\Sigma_\ast$ and $\Sigma_g$ being measured in $\msun~{\rm yr^{-1}}~{\rm kpc}^{-2}$ and $\msun~{\rm kpc}^{-2}$, correspondingly.) the observed star formation rate (SFR) of 25 M$_\odot$ yr$^{-1}$ within the half-light radius $R_h\sim 0.6\pm 0.3$ kpc {implies} a total gas {mass $M_g\approx 3\times 10^8\msun$}. 

A SFR of $\dot M_\star\sim 25~\msun$ yr$^{-1}$ with a Salpeter initial mass function (IMF) produces supernovae (SNe) explosions at a rate $\nu_{\rm SN} \sim 0.7$ yr$^{-1}$ (with $\nu_m = 130^{-1}~\msun^{-1}$ being the specific SNe rate for a Salpeter IMF), each injecting 
a fraction $\mu_Z\sim 0.2$ of the pre-SN mass 
in metals, and a considerable fraction ($0.4$) of them 
in form of dust \citep{Todini2001,Bianchi2007,Matsuura2011,Gall2014,Lau2015}. 
Direct measurement of dust in $z=6\hbox{--}8.5$ galaxies lead \citet{Lesniewska2019} to an estimate of the ejected dust mass, as $\mu_d\sim 0.1\hbox{--} 1~\msun$ per SN. Observations of cold dust in Cas A allowed \citet{Krause2004} to set an upper limit $\mu_d\le 0.2~\msun$. A more conservative estimate, $\mu_d \sim 0.1~\msun$ follows from simulations by \citet{Todini2001}. Here, we use a conservative value of $\mu_d\sim 0.1\, M_\odot$.  

With these assumptions, the total estimated mass of metals and dust injected into the ISM within 40 Myr by SNe is $M_Z\sim 8\times 10^6~\msun$, and $M_d\sim 8\times 10^5~\msun$, respectively. The dust-to-gas mass ratio is, therefore,  $\zeta_d\sim 3\times 10^{-3}$, nearly $30$\% of the Milky Way value. Assuming the dust to be  Milky Way-type, the expected optical depth at UV ($1000$ \AA) $\tau_{uv}\sim \sigma_{uv}nR\sim 10^{-21}\times 3\times 10^{22}\zeta_d\zeta_{d,MW}^{-1}\sim 10$, where $\sigma_{uv}\sim 10^{-21}$ cm$^2$  \citep[][]{Draine84,Corrales2016}, and $nR=\Sigma_g/2\mu m_{\rm H}$, $\zeta_{d,MW}\approx 120^{-1}$. This estimate strongly conflicts with the observational transperancy of the ISM in GN-z11: $E(B-V)\sim 0.01\pm 0.01$  \citep{Jiang2020}. 

Possible explanations can be sought either from the suppression of dust yield by reverse shocks from SNe, or with destruction of dust by SNe triggered shocks, and/or with evacuation of dust by gas outflows driven by stellar activity, or a combination of all these.   

\subsection{Reducing dust attenuation}\label{dd} 

In the circumstance under consideration here, dust grains are mostly destroyed by thermal sputtering\footnote{We do not account nonthermal sputtering due to betatron acceleration and kinetic sputtering of dust particles moving in ambient gas supersonically. This corresponds to a lower estimate of dust destruction.}. Destruction of dust in similar conditions has been taken into account previously by \citet{Arata2019}, by scaling  the dust-to-gas ratio in proportion to the fraction of HI in the hot gas. Here, we use a different approach, and implement dust destruction depending on the dust radius $a$ by comparing the sputtering time $t_{sp}(a)$ with the dynamical time of the cell. More explicitly, we do not remove all dust particles entering into the volume occupied by hot gas, but only when 
\blue{the sputtering time is shorter than the dynamical time}. 
The characteristic sputtering time $t_{sp}$ and the gas cooling time $t_{cool}$ at temperatures close to $T\sim 10^6$ K: $t_{sp}\sim 3\times 10^5T_6^{-2.5}a_{0.1}n^{-1}$ yr at $T\simlt 10^6$ K, and $t_{sp}\sim 10^5T_6^{-3}a_{0.1}n^{-1}$ yr for $T\ge 10^6$ K  
\citep{Draine1979,Draine2011}, $t_{cool}\sim 2\times 10^4T_6^{1.5}n^{-1}$ yr \citep{v13}, $a_{0.1}=a/0.1~\mu$m, $T_6=T/10^6$ K. Therefore, dust particles with radii 
$a>100$ \AA  $\,$can partially survive at temperatures within the hot bubble formed by a starburst with SNe shock velocities of $v_s\sim 100$ km s$^{-1}$.    
We assume, in the {\it unperturbed} medium, a given initial dust-size spectrum $n_d(a,t=0)$, and then we compare in each computational cell {\it behind} the shock front the characteristic dust thermal sputtering time and the dynamical time $t_D$. 
When $t_{sp}(a)/t_D\leq 1$, dust particles of a given size $a$ are removed from the distribution function $n_d(a, t=0)$ within the cell.

Dust can also be destroyed {\it in situ} during the process of delivery by an expanding SN ejecta because of the reverse shock \citep{Bianchi2007,Nozawa2007}, considerbaly reducing the fraction of small size dust particles \citep[$a\simlt 10^{-3}\mu$m, see][lower panel in their Fig. 6]{Nozawa2007}.  In the present case this effect can be important only at very initial stages of star formation in the central stellar clusters, when the remnants from individual SNe expand into unperturbed surrouding gas in a cluster. After the merging of remnants from the very first 1000 SNe, they form a joint superbubble with a hot low-density gas inside (see in Sec. \ref{sec:num}). In these conditions reverse shocks do not form and dust particles can survive  {\it in situ} destruction. In order to examine how the shape of the {\it unperturbed} dust-size spectrum  affects the results, we consider the two models: one with the ``standard'' MRN spectrum $n_d(a) \propto a^{-3.5}$ \citep[][]{Mathis1977}, and the other with a ``flat'' spectrum in the form $n_d(a) \propto a^{-1.5}$ similar to that used by \citet{Maiolino2004} and confirmed  numerically by \citet{Nozawa2007} for conditions in early galaxies. 

Since the optical depth scales as $\tau_{uv}\propto nR\propto R^{-2}$ for a given dust mass kept, it can drop to the detection level $E(B-V)\sim 0.01$ if the dust cloud expands up to $R\simgt 10$ kpc. Attributing such an expansion to SNe explosions, with a rate of $\nu_{\rm SN}\sim 0.2$ yr$^{-1}$, corresponding to mechanical luminosity $L_{\rm sn}\sim 6 \times 10^{42}$ erg s$^{-1}$, the bubble size evolution is given by,
$
R(t)\sim (\eta L_{\rm sn}t^3)^{1/5} \, \rho^{-1/5}\,,
$  
\noindent 
where $\rho$ is the density of ambient ISM gas. 
For $\rho_{100}=\rho \,/ (100\, m_{\rm H})$ g cm$^{-3}$, and $\eta= 0.5\eta_{0.5}$, a radiative correction factor, the time taken for the shell to reach $10$ kpc, is
$
t \approx 7 \, {\rm Myr} \,  (R / 10 \, {\rm kpc})^{5/3} \, \rho_{100}^{1/3} \, \eta_{0.5}^{-1/3}\,,
$
\noindent
which is shorter than the star formation time scale of GN-z11.
 
In order to confirm these estimates we simulate SNe driven outflow assuming the following model for a galaxy before explosion.    
 
\section{Model setup}\label{mod} 
We carry out 3-D hydrodynamic simulations (Cartesian geometry) of multiple SNe explosions in the central part of a galaxy. The code is based on the unsplit total variation diminishing (TVD) approach that provides high-resolution capturing of shocks and prevents unphysical oscillations. We have implemented the Monotonic Upstream-Centered Scheme for Conservation Laws (MUSCL)-Hancock scheme and the Haarten-Lax-van Leer-Contact (HLLC) method \citep[see e.g.][]{Toro1999} as an approximate Riemann solver. This code has successfully passed the whole set of tests proposed in \citet{Klingenberg2007}. It is difficult to pin down the exact geometry, size and profiles of gaseous and the stellar content of galaxy at such a high redshift. At the same time, one has to fix these in the initial set up in order to simulate. Our idea in the following exercise is to have a simple yet robust set up so that does not mitigate the essential physics. One important problem one faces in this regard is that of gas overcooling in the central region, if one assumes gas to be distributed spherically. We avoid this by using a disc geometry as recently found in $z\simeq 9-12$ galaxies \citep{Adams2022}, although the physical processes we attempt to demonstrate here robust enough, as the above estimates show, and do not crucially depend on these particular assumptions. We consider gas in the central region to be initially in the hydrostatic equilibrium in the gravitational potential of a dark matter halo, and a stellar disc. 

The dark matter halo is assumed to follow the Navarro-Frenk-White (NFW) profile, with virial radius $12.4$ kpc with a concentration parameter $c=6$, corresponding to a total galactic mass of $10^{11} \msun$ at $z=11$. We use a softening length of $d=3$ kpc for the NFW potential. The total amount of gas within the virial radius is assumed to be a cosmic fraction ($16\%$) of the total mass. The disc potential is provided by a total mass of $10^9 \msun$, which can be thought of as comprising gas and stars, recalling that the observed stellar mass is $1.3\pm 0.6 \times 10^9 \msun$. We use the Miyamoto-Nagai disc potential with scalelength $a=2$ kpc and scaleheight $b=0.8$ kpc \citep{Miyamoto1975}.  We are interested in the vertical (relative to a disc) motions of a gas, because this direction is favourable for observing a naked stellar population (i.e. low extinction similar to that measured in the observations). We therefore consider a vertically stratified medium, with a mid-plane density of $100$ cm$^{-3}$ and a scale height of $180$ pc. 
 
In order to simulate the effects of energy, metal and dust injection by SNe as specified in Sec. \ref{sec:optdept}, we set up a stellar population within star clusters randomly spread in an ellipsoidal region with major and minor axes of $0.7$ and  $0.25$ kpc. The masses of star clusters range randomly between $M_{cl} \sim 5\times 10^4 - 10^6~\msun$ according to a power-law distribution: $N\sim M_{cl}^{-2}$ \citep{Krumholz2019}. Each cluster includes $N_p=\nu_m M_{cl}$ massive stars, which are progenitors of SNe, with masses ranging between $M\sim 10-40~\msun$.  
We also assume a burst-like star formation model for all the stellar clusters inside the ellipsoid, with the SNe spread randomly inside the corresponding cluster. 
The lifetime of massive stars relates stellar mass as 
$t_l \propto M^{-1.57}$ \citep{Iben2012}. The (longest) lifetime of $10~\msun$ stars is $t_l\sim 15$~Myr. We begin our run when the most massive SN explodes.

To reduce the computation time, we assume that 200 SNe with equal mass of their progenitors explode simultaneously as a single SN, and accordingly each of such explosions inject jointly  $N_{sn}$ times the energy, mass and metals of a single {explosion}. The energy of such {explosion} is $200E_{\rm B}=2\times 10^{53}$~erg, which is injected in form of thermal energy. The injected masses of gas and metals are also increased by a factor $N_{sn} = 200$, i.e. the mass load for a {joint explosion} is $\sim 1600-8000~\msun$ ($15\times200 = 3000~\msun$ on average over the IMF) correspondingly. The total number of {such explosions occurring} during $\Delta t_\ast \sim \langle t_l\rangle\sim 15$~Myr is $N_h\sim 5700$. Therefore, the total ejected energy is $\sim  10^{57}$~erg, which, over $15$ Myr of SNe explosions (see above), corresponds to a power of $2.5 \times 10^{42}$ erg s$^{-1}$, as required. The corresponding mass ejection is ${\cal M}_{SN} \sim 3000~\msun\times 5700\sim 1.7\times 10^7\msun$. 

{Radiative cooling is implemented in the simulations with a tabulated non-equilibrium cooling function $\Lambda(T,Z)$ calculated as described in details in \citep{v13}, and splined accordingly for $T_i$ and $Z_i$ at different time steps $t_i$. The functions are obtained for isochoric cooling from $10^8$~K down to 10~K for 68 discret values of metallicity within the range [Z/H]$=-4..1$. The simulations are performed with a physical cell size of 10~pc.}
This size is sufficient to resolve the evolution of a single super-SN with energy $200E_{\rm B}=2\times 10^{53}$~erg, as mentioned above \footnote{The cooling time is $t_c \sim k_B T/n\Lambda \sim 3\times 10^{12}$~s, so that for the typical shock velocity $v\sim 200$~km/s the SN shell at the beginning of radiative phase is greater than the cell size}. Note that after $\sim 0.5$~Myr from initiating the starburst in the galaxy, the majority of shells formed around stellar clusters begin to interact with each other and merge into a combined bubble.


\section{Results}\label{sec:num} 
The disturbance in the ISM due to combined SNe proceeds to engulf the galaxy; the `breakthrough' from the disk in the vertical direction happens at $\sim 5$ Myr, and the whole gaseous disk is enveloped in $\sim 16$ Myr. Thereafter the shock wave proceeds through the circumgalactic medium with a roughly spherical shape. The average temperature within the shock wave is $\approx 10^7$ K, although radiative cooling 
leads to cloud formation with dense patches embedded in hot and tenuous medium.

\begin{figure}
\center
\includegraphics[width=8cm]{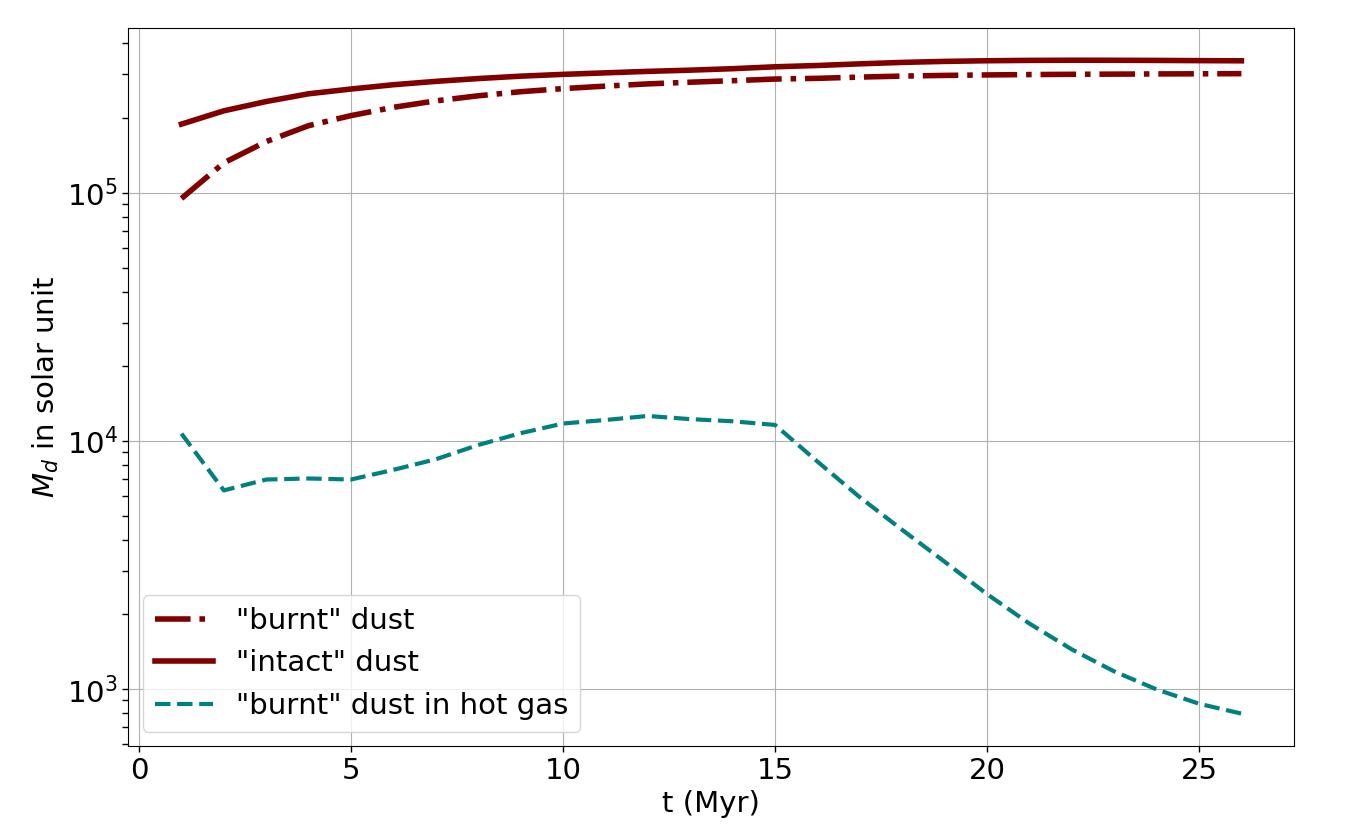}
\caption{ 
Evolution of dust mass. Solid line shows the history of the total dust mass injected into the bubble by SNe, and the dash-dotted line depicts the mass of dust grains that have been processed by hot gas and survived sputtering.  
The dust mass that have suffered sputtering in the hotest parts with $T\geq 10^6$ K is presented by dashed line.} 
\label{sput}
\end{figure}

As mentioned above, SNe explosions enrich the galaxy ISM during the evolution until the least massive $M=10~\msun$ progenitors exhaust. In total, around $2\times 10^6~\msun$ of metals and $\sim 8\times 10^5~\msun$ (for the fraction of metals in solid phase $\xi_d \sim 0.4Z$) of dust are supplied into the superbubble as shown in Figure \ref{sput}, where the total mass of dust (solid lines) and the mass of dust experienced sputtering (dash-dotted line), along with the dust locked in the hot ($T\geq 10^6$ K) superbubble interior indicated by dashed line are shown. It is seen that although the mass fraction of dust within the hot region during the first $1\hbox{--}5$ Myr can be as high as $\sim 0.03-0.1$, it drops  below $\sim 10^{-2}$ afterwards, while the rest accumulates in the cold clumps.

\begin{figure*}
\center
\includegraphics[width=17cm]{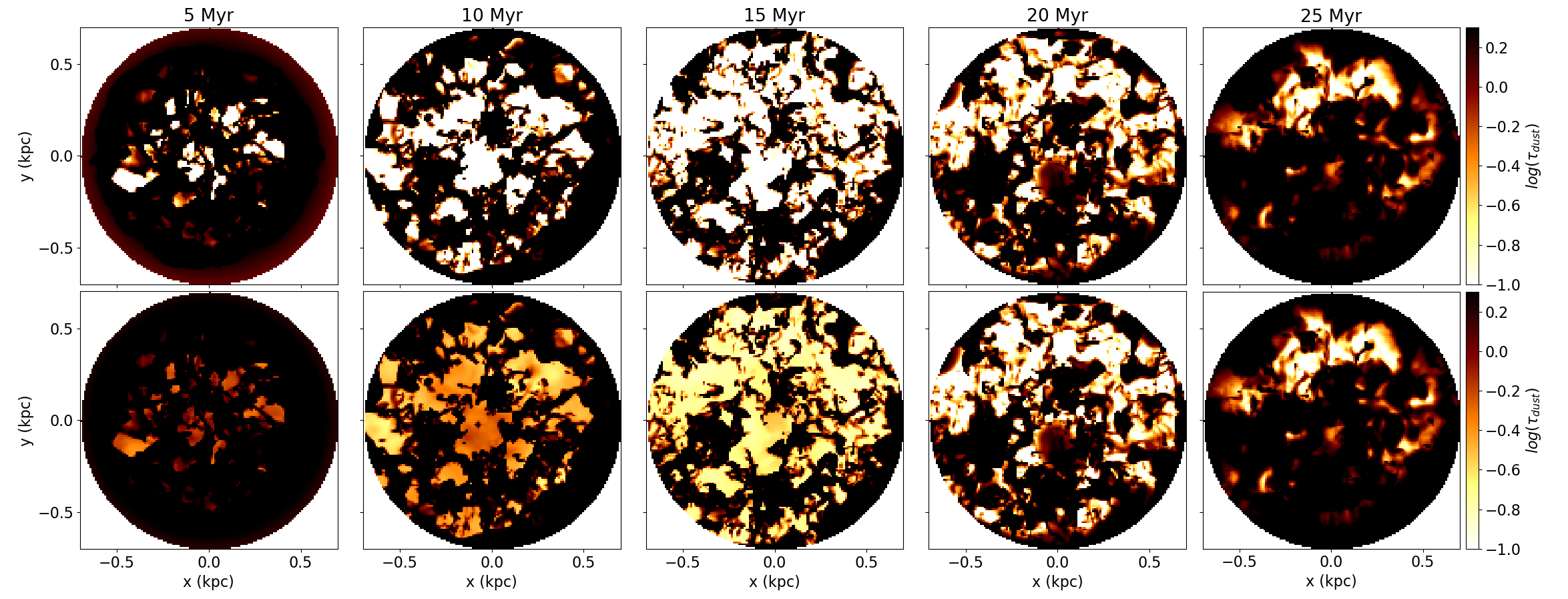}
\caption{
Maps from left to right show the evolution of a patchy extinction distribution over the bubble's front-side hemisphere: the upper row illustrates the distribution when dust sputtering is taken into account, while the lower panel shows the case without dust destruction. Clearly seen in both cases are pervading spots of a lowered extinction grown under Rayleigh-Taylor instability. 
}
\label{patch}
\end{figure*}

Since the gas density in the hot interior drops below $10^{-2}$ cm$^{-3}$ by $\sim 20$ Myr, the dust sputtering time is increased ($t_{sp}\sim 10 a_{0.1}$ Myr). Also, most of the dust is pushed into the conducive environment of the dense and cooler shell where it can survive. In other words, dust destruction is important only in the initial period of bubble growth ($\le 5$ Myr) when the shock is confined within a few scale heights.

\subsection{Extinction of the outflow} \label{sec:extinct}
\subsubsection{Fog clearing} \label{dust_mod}
Figure \ref{patch} shows simulated maps of optical depth at different epochs, as seen face-on towards the hemisphere facing the observer (hereafter referred to as `front-side hemisphere') of the growing bubble, with (upper row) and without (lower row) sputtering effects. 
Mapping of optical depth of one hemisphere seems to be sufficient to demonstrate whether or not the radiation from stars, located mostly in the disc, would be visible.
Left to right panels illustrate the bubble ages from 5 to 25 Myr as labelled. In both cases, we find pervading areas of a lowered extinction that originates from Rayleigh-Taylor instability of expanding but decelerating shell. Note that, at the epoch of maximum transperancy ($\approx 15$ Myr), even the clear patches in the lower panel (without dust sputtering) are darker in colour than in the upper panel (with dust sputtering). This implies that dust sputtering increases the overall transparency. Although this is intuitively expected, our simulation demonstrates it quantitively. It is seen, however, that in spite of dust destruction, a fraction of the
bubble area remains dim because of partially survived dust.  

\begin{figure*}
\includegraphics[width=16cm]{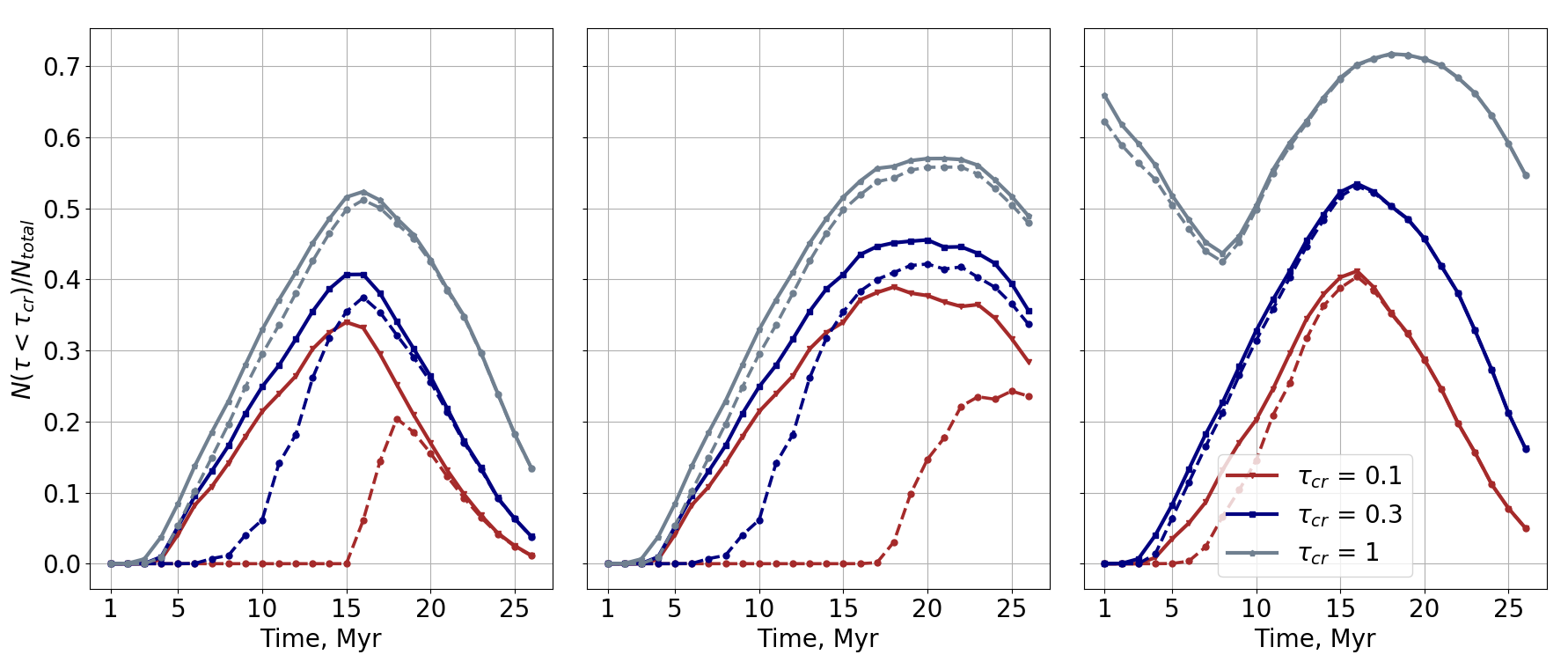} 
\caption{
Left panel shows the evolution of the fraction of pixels $f_{\tau_{uv}}(t)$ of the front-side hemisphere with optical depth lower than a given level, as shown in the legend, for the model with the IMF truncated at $M_{{\rm SN},{\rm min}}=10~\msun$ (when the SN explosion stops at $t\simeq 15$ Myr). Comparison of the $f_{\tau}(t)$ for models with the intact (dashed curve) and `burnt' (solid curves) dust  shows that the shell is more transparent in the latter case. Middle panel shows the model with $M_{{\rm SN},{\rm min}}=8~\msun$, for which SNe explosions cease at $t\simeq 25$ Myr. Right panel: same as the left panel ($M_{{\rm SN},{\rm min}}=10~\msun$) for the dust with a `flat' size distribution $n(a)\propto a^{-1.5}$ and a minimal grain radius of $a_{\rm min}=0.003~\mu$m as in \citep[][Fig. 6, lower panel]{Nozawa2007}. An obvious difference for the `flat' dust is its higher transparancy -- at the very initial stages of the bubble evolution a considerable fraction of the front-side the hemisphere is `half'-transparent, $f_{\tau_{cr}=1}(t)\sim 0.6$. At later times, $f_{\tau_{cr}=1}(t)$ decreases while the SNe dust supply compete its desctruction by shocks at $t\sim 8$ Myrs.
 }
\label{transp}
\end{figure*}

Figure \ref{transp} illustrates cleaning in the expanding shell quantitatively -- it represents the fraction of area of the front-side shell with optical depth $\tau_{uv}$ lower than a certain level as shown in the legend. Initially (although not shown here) one has $\tau_{uv}>1$ and $f_{\tau_{uv}<1}(t)\simeq 0$ as depicted by the dashed curve. At later epochs, the shell fragments under Rayleigh-Taylor instability and forms relatively transparent windows, such that $f_{\tau_{uv}<1}$ cannot be neglected anymore, particularly in the model that has sputtered (`burnt') dust. Left panel in Fig. \ref{transp} represent the model with threshold SN progenitor mass of $10~\msun$. 
The difference between the solid (`burnt') and dashed (`intact') curves is large in the case of the (red) curve for $\tau_{uv}<0.1$, which represents the clearest patches. This difference is large in the epochs of continuing SNe explosions (before 15 Myr), and the gap decreases at later times, due to the decreasing influence of SNe shocks.

Expected variations of dust properties (e.g., dust-size distribution) in early galaxies at  $z\simgt 6$ as compared to local galaxies can make their appearance distinctly different \citep{Nozawa2007}. To study this, 
we also calculated the optical depth evolution of the expanding dusty outflows with a `flat' dust size distribution $n_d(a)\propto a^{-1.5}$, 
normalized by the dust mass in the galaxy and in the ejecta. Due to a deficit of small dust particles compared to the `standard' MRN dust size distribution \citep[
][]{Mathis1977} $n_d(a)\propto a^{-3.5}$, the extinction $A_V$ in this model is lower by factor $\sim (3/5)(a_M/a_m)^{0.5}$, $a_m$ and $a_M$ being the radii of the smallest and the largest grains, assuming the same dust-to-gas mass ratio, in both cases we assumed $a_m=30$\AA, $a_M=0.15~\mu$m. The consequences are seen in the right panel of Fig. \ref{transp}: during the evolution the bubble remains more transparent as compared with the middle panel. Moreover, the difference between the fractions of hemisphere with optical depth below a certain level in models with the `burnt' and intact grains are comparable, particularly for curves $\tau_{cr}=0.1$ in the middle and right panels.  

\subsubsection{From clearing back to darkening}
The fraction of transparent areas monotonously increases until $t= 15$ Myr, when the SNe cease to explode. Soon after the sputtering from shock waves slows down,  the pattern of transparancy maps in the lower (intact dust) and upper (burnt dust) rows in Fig. \ref{patch} approach each other and become nearly identical. This is also seen in Fig. \ref{transp}, and is connected with decaying turbulent motions within the superbubble that redistribute the remaining grains, and mix the regions with low and high extinction, resulting in diminishing the fraction of the transparent area. The time required for this mixing is $t_{mix}\sim R_{bubble}/v_t\sim 7$ Myr (for a bubble radius of $0.7$ kpc, and turbulent velocity $v_t\sim 100$ km s$^{-1}$), and is consistent with the sequence shown in Fig. \ref{patch}. 

Stellar population with 
threshold SN progenitor mass of
$8~\msun$ shows a longer transition from cleaning to darkening stage, $t\simeq 25$ Myr (middle panel of Fig. \ref{transp}). This is connected with a longer process of fragmenting the ouflow due to Rayleigh-Taylor instability and dust sputtering in hot regions.   

It is clearly seen in Figure \ref{transp} that starting from $t=15$ Myr, the areas of $\tau_{uv}<0.1$, with the corresponding reddening of $E(B-V)\sim 0.03$ 
for the standard (Milky Way) extinction law  with $R_V\simeq 3$ \citep[see in ][]{Draine84,Weingartner2001,Draine2011}, occupy a fraction $\sim 0.3$ in the model with $M_{{\rm 
SN},{\rm min}}=10~\msun$. When dust sputtering is taken into account, the parameter $R_V$ increases to the value typical for AGNs $R_V\simeq 5$ \citep[][]
{Maiolino2001,Gaskell2004}, and the reddening approaches the value measured in GN-z11 $E(B-V)\simlt 0.02$ \citep{Jiang2020}. For the flat dust-size distribution, 
the parameter $R_V$ increases from $R_V\simeq 6.1$ before entering shock, to $R_V\simeq 6.2$ after shock processing. This diminishes the reddening to the value measured in GN-z11 $E(B-V)\simlt 0.016$.

At latter stages, at $t=20$ Myr a small part of the area becomes as transparent as $\tau_{uv}\sim 0.01$, however regions with $\tau_{uv}\simlt 0.1$ shrinks below $\simlt$ 15\%. At later stages $t=25$ Myr nearly 50\% of the shell becomes covered with optical depth upto $\tau_{uv}\simgt 3$ in accordance with the maps shown in Fig. \ref{patch}.  However, the model with truncated IMF at $M_{{\rm SN},{\rm min}}=8~\msun$ with SNe acting on longer times $t\simeq 25$ Myr, shows even a higher fraction of $\simeq 0.4$ of optically thin area with $\tau_{uv}< 0.1$ for as long as 23 Myr, and in later 5 to 7 Myr, they dim quickly with the fraction of transparent areas $\tau_{uv}< 0.1$ below $\simeq$ 15\% at $t>28$ Myr.


\section{Discussion} 
Even though the amount of dust that can be produced and ejected by SNe during the entire period of star formation in GN-z11 ($\sim 40$ Myr) is quite sufficient  ($M_d\sim 8\times 10^5~\msun$) to enshroud the stellar light, the mechanical power from the same SNe is capable to blow the dust shell out and to break in it patchy transparent windows. The apparent `puzzling' lack of dust in the galaxy GN-z11 might reflect that detection of this galaxy was rather accidental, when its evolutionary phase passed through a period of `transparency'. Switching to infrared wavelengths may not help, because dust heating for grains embedded in hot gas is not substantial enough to raise their temperature much above that of CMB: the dust temperature estimated by the peak in its emission spectrum is $T_d\simeq 34$ K, with the peak flux $F_{70~\mu{\rm m}}\simeq 0.7$ mJy, nearly the detection threshold of B-band in ALMA. This takes into account the fact that  dust emission will be embedded in the CMB \citep[see discussion in ][]{Dacunha2013}.  

As the metal enriched and dusty shell settles down to the galactic disc and stellar population  gets at least partially obscured, the galaxy becomes veiled, and therefore, avoids detection in rest-frame UV and optical range. At the same time, the measured stellar luminosity in GN-z11 $L_\ast\sim 2\times 10^{10}L_\odot$ can heat the settled dust only up to temperatures that are a few per cents higher than the CMB $T_{cmb}\simeq 32.4$ K at $z=11$. For this reason, the galaxy cannot be observed in the infrared as well. Following these arguments and accounting for a relatively short period of the starburst in GN-z11 --- 10 \% of the Universe age at $z=11$ --- one may assume that a considerable fraction of galaxies similar to those determining the actual value of the galaxy UV luminosity function (GUVLF) can be hidden beneath the dust -- a factor of 10 higher than currently inferred. Such a possibility has been recently confirmed from the Early Release Science Program CEERS that identified a set of 33 heavily obscured $A_v\sim 2$ massive star-forming galaxies at $z=2-8$, which are not detected in HST surveys \citep{Barrufet2022}.  

Recent observations of galaxies at the beginning of cosmic dawn and measurements of the GUVLF at 
$z\simgt 8-10$ \citep{Rojas2020} show 
 outliers at the bright end of the GUVLF. We speculate that 
they may show periods of starburst `transparency' during their evolution, and their relative fraction compared to those lying in the `standard' luminosity function at these $z$, 
is proportional to the duty cycle of the `transparent' phase. Note that the transparent phase can 
continue longer for a flat dust-size distribution, 
as seen in the right panel in Fig. \ref{transp}.    

 Recently {\it James Webb Space Telescope} (JWST) found surprisingly abundant massive and bright blue $M_{UV}<-23$ galaxies at $z>11$ \citep{Windhorst2022,Yan2022}.  One possible explanation is that $z>10$ galaxies mostly had high star formation efficiency, and were therefore brighter in UV than latter galaxies \citep{Castellano2022,Mason2022}. Other scenarios have been described by \citet{Ziparo2022}. One of them suggests radiation driven dust ejection out of galaxies, while another model assumes dust to be segregated in dense compact clouds with a low covering factor. \cite{Ziparo2022} have stated that the recently reported non-detection of the rest-frame [OIII]88$\mu$m line in the $z\simeq 13$ galaxy GNZ2/GL-z13 \citep{Bakx2022, Popping2022} can be explained by the first scenario, since the radiatively driven dust outflow will also evacuate gas, and therefore, OIII ions. In our model,  
 the absence of OIII ions can also be easily explained, because
 the optically thin regions of the expanding gas are filled with
 diffuse and highly ionized hot gas $T> 3\times 10^5$ K. These regions contain a very small fraction of OIII ions, $f_{\rm OIII}\simlt 0.01$ \citep[as discussed in ][ their Fig. 2.1b, upper panel]{Gnat2007}. Threfore, the observed lack of OIII is also a generic prediction of SN shock driven outflowing gas.     


\section{Summary} \label{sec:sum}

Our study of the dynamical and observational consequences of the high-$z$ starburst galaxy GN-z11 demonstrates that: 
\begin{enumerate} 
\item SFR  measured in GN-z11 is sufficient to remove dust beyond the galactic disc to heights up to $\sim 1.5$ kpc, thereby making it transparent. Dust with a flat size distribution is potentially more conducive for making the disc transparent. 
\item Within $\sim$ 10 Myr after the end of SNe explosions, the evacuated material along with dust fall back and obscure stellar light until conditions for a next starburst are met with. 
\item Such an obscuration may hide many starburst galaxies avoiding detection. 
\item The total amount of dust is insufficient to be seen in the rest-frame FIR/submm domain, making GN-z11-like galaxies unobservable in these wavebands. 
\end{enumerate} 

\vspace{1cm}
We thank Andrea Ferrara for helpful discussion and an anonymous referee for detailed comments.
The work of SD is done under partial support from the project ``New Scientific Groups LPI'' 41-2020.

\section*{Data Availability}

The data underlying this article are available in the article.


\end{document}